\documentclass[10pt,letterpaper]{article}
\usepackage{opex3}
\usepackage{hangcaption}
\usepackage{array}
\usepackage{indentfirst} 
\usepackage{amsmath}
\usepackage{amssymb}

\newcommand{\nufourier}[1]{NUF \biggl[#1 \biggl]}
\newcommand{\snufourier}[1]{NUF \bigl[#1 \bigr]}
\newcommand{\x}[1]{{\bf x}_{#1}}

\begin{document}




\title{Arbitrary shape surface Fresnel diffraction}

\author{
Tomoyoshi Shimobaba,$^{1,*}$ 
Nobuyuki Masuda,$^1$
and
Tomoyoshi Ito$^1$
}

\address{$^1$ Graduate School of Engineering, Chiba University, 1-33 Yayoi-cho, Inage-ku, Chiba 263-8522, Japan}
\email{$^*$ shimobaba@faculty.chiba-u.jp}


\begin{abstract} 
Fresnel diffraction calculation on an arbitrary shape surface is proposed.
This method is capable of calculating Fresnel diffraction from a source surface with an arbitrary shape to a planar destination surface.
Although such calculation can be readily calculated by the direct integral of a diffraction calculation, the calculation cost is proportional to $O(N^2)$ in one dimensional or $O(N^4)$ in two dimensional cases, where $N$ is the number of sampling points.
However, the calculation cost of the proposed method is $O(N \log N)$ in one dimensional or $O(N^2 \log N)$ in two dimensional cases using non-uniform fast Fourier transform.
\end{abstract}

\ocis{(0070) Fourier optics and signal processing; (090.1760) Computer holography; (090.2870) Holographic display; (090.5694) Real-time holography; (090.1995) Digital holography.} 

\


\section{Introduction}
\noindent Diffraction calculations such as Huygens diffraction, Fresnel diffraction and angular spectrum method, are important tools in wide-ranging optics \cite{goodman, okan}, ultrasonic \cite{ultrasonic}, X-ray \cite{xray}  and so forth.
Its applications in optics include computer-generated-hologram (CGH) and digital holography \cite{poon}, phase retrieval, image encryption and decryption and so forth.

Fast Fourier transform (FFT)-based diffraction calculations according to their convolution or Fourier transform form are used in these applications.
The FFT-based diffraction calculations, however, can only be applied to planar surfaces in parallel.
In order to apply the methods to a non-parallel planar surface, many methods have been proposed: for example, non-parallel Fresnel diffractions \cite{Frere,Frere2,Yu,sakamoto} and non-parallel angular spectrum methods \cite{tomassi,matsu,onural,Ahrenberg}.
Unfortunately, these non-parallel diffractions are limited to planar surfaces.
If we calculate a source surface with arbitrary shape using non-parallel diffractions, we need to approximate the arbitrary shape surface with many small non-parallel planar surfaces, and then, we need to calculate the non-parallel diffraction per the small non-parallel planar surfaces.

In this paper, we propose Fresnel diffraction calculation on an arbitrary shape surface.
Without approximating an arbitrary shape surface with small non-parallel planar surfaces, this method is capable of calculating Fresnel diffraction from a source arbitrary shape surface to a planar destination surface.
Although such calculation can be readily calculated by the direct integral of a diffraction calculation, the calculation cost is proportional to $O(N^2)$ in one dimensional or $O(N^4)$ in two dimensional cases, where $N$ is the number of sampling points.
However, the calculation cost of the proposed method is $O(N \log N)$ in one dimensional or $O(N^2 \log N)$ in two dimensional cases using non-uniform fast Fourier transform.

In Section 2, we describe the arbitrary shape surface Fresnel diffraction.
In Section 3, we present the numerical results.
Section 4 concludes this work.

\section{Arbitrary shape surface Fresnel diffraction}

Let us begin with Huygens diffraction.
Huygens diffraction \cite{goodman} on a planar surface is expressed as:
\begin{equation}
\begin{aligned}
u_2({\bf x_2})= \frac{z_0}{i \lambda} \int \!\! \int  u_I(\x{1}) u_1({\bf x_1})
\frac{\exp(i k r)}{r^2} d{\bf x_1},
\end{aligned}
\label{eqn:huy_diff}
\end{equation}
where, $u_1(\x{1})$ and $u_2(\x{2})$ are planar source and destination surfaces, $\x{1}$ and $\x{2}$ are the position vectors on the source and destination surfaces,  $\lambda$ and $k$ are the wavelength and wave number of light, and $r=\sqrt{|{\bf x_2} - {\bf x_1}|^2 + z_0^2 }$, where $z_0$ is the distance between the source and destination surfaces.
$u_I(\x{1})$ is the incident wave to the source surface. 
If the incident wave is used as a planar wave that is perpendicular to the optical axis, we can treat as $u_I(\x{1})=1$.

\begin{figure}[htb]
\centerline{
\includegraphics[width=10cm]{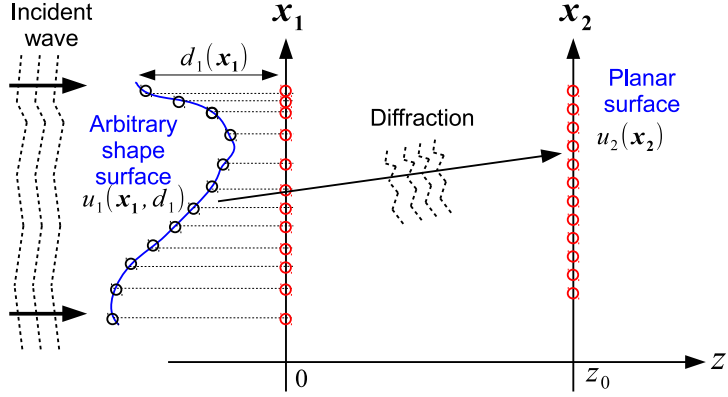}}
 \caption{Diffraction calculation between source surface with arbitrary and planar destination surface.}
\label{fig:system}
\end{figure}

We expand the Huygens diffraction to a source surface with arbitrary shape.
As shown in Fig. \ref{fig:system}, the source surface with arbitrary shape $u_1({\bf x_1},d_1)$ is defined by the displacement $d_1=d_1(\x{1})$ at the position $\x{1}$.
Note that when the arbitrary shape surface is uniform-sampled, the corresponding coordinate $\x{1}$ is non-uniform-sampled depending on the slope of $u_1({\bf x_1},d_1)$ at the position $\x{1}$.
Huygens diffraction on an arbitrary shape surface is expressed by:
\begin{equation}
\begin{aligned}
u({\bf x_2})= \frac{z_0}{i \lambda} \int \!\! \int u_I(\x{1}, d_1) u_1({\bf x_1},d_1)
\frac{\exp(i k r)}{r^2} d{\bf x_1},
\end{aligned}
\label{eqn:huy_diff2}
\end{equation}

\begin{equation}
\begin{aligned}
r= \sqrt{|{\bf x_2} - {\bf x_1}|^2 + (z_0  - d_1)^2 }.
\end{aligned}
\label{eqn:r1}
\end{equation}

Here, if the incident wave $u_I(\x{1}, d_1)$ is used as a planar wave that is perpendicular to the optical axis, the incident wave is expressed as $u_I(\x{1}, d_1)=\exp(i k d_1)$.

Applying Fresnel approximation to Eq.(\ref{eqn:r1}) using $r_0=z_0  - d_1$, we can obtain the following approximation:
\begin{equation}
\begin{aligned}
r \approx r_0+\frac{\x{1}^2}{2 r_0} - \frac{\x{1}\x{2}}{r_0}+\frac{\x{2}^2}{2 r_0}.
\end{aligned}
\label{eqn:r2}
\end{equation}

And, we also approximate $r^2 \approx z_0^2$ in the integral of Eq.(\ref{eqn:huy_diff2}).
Therefore, we obtain Fresnel diffraction on an arbitray surface:
\begin{equation}
\begin{aligned}
u({\bf x_2})&= 
\frac{\exp(i k z_0)}{i \lambda z_0} 
\int \!\! \int 
u_I(\x{1},d_1) u_1(\x{1},d_1) \\
&	\exp(ik(-d_1 + \frac{\x{1}^2}{2 r_0}))  
	\exp(- ik \frac{ \x{1}\x{2}}{ r_0}) 
	\exp(\frac{\x{2}^2}{2 r_0})	d{\bf x_1}
\end{aligned}
\label{eqn:fre_diff1}
\end{equation}

The above Eqs.(\ref{eqn:huy_diff2}) and (\ref{eqn:fre_diff1}) can be treated an arbitrary shape surface.
We can readily calculate by the direct integral with regard to these equations; however, the calculation cost is $O(N^2)$ in one dimensional or $O(N^4)$ in two dimensional cases, where $N$ is the number of sampling points, because we cannot calculate them using Fourier transform.

In order to obtain the Fourier form of Eq.(\ref{eqn:fre_diff1}), we approximate the third exponential term in the integration as follows:

\begin{equation}
\begin{aligned}
\exp(ik \frac{\x{2}^2}{2 r_0}) \approx \exp(ik \frac{\x{2}^2}{2 z_0}).
\end{aligned}
\label{eqn:fre_diff4}
\end{equation}

Eventually, we obtain the following equation:
\begin{equation}
\begin{aligned}
u({\bf x_2})&= 
\frac{\exp(i k (z_0  + \frac{\x{2}^2}{2 z_0}))}{i \lambda z_0} \\
&	\int \!\! \int 
 	u_I(\x{1},d_1) u_1(\x{1},d_1)
	\exp(ik(-d_1 + \frac{\x{1}^2}{2 (z_0 - d_1)})) \\
&	\exp(ik \frac{\x{1}\x{2}}{z_0 - d_1}) 
	d{\bf x_1}.
\end{aligned}
\label{eqn:fre_diff5}
\end{equation}

Because the coordinate $\x{1}$ is sampled by the non-uniform sampling rates, instead of (uniform) Fourier transform,  
we can calculate the above equation using non-uniform Fourier transform (NUFT):
\begin{equation}
\begin{aligned}
& u(\x{2})= 
\frac{\exp(i k (z_0 + \frac{\x{2}^2}{2 z_0}))}{i \lambda z_0} \\
&	\nufourier{
 	u_I(\x{1},d_1) u_1(\x{1},d_1)
	\exp(ik(-d_1 + \frac{\x{1}^2}{2 (z_0 - d_1)}))},
\end{aligned}
\label{eqn:fre_diff_nuf}
\end{equation}
where, $\snufourier{\cdot}$ denotes NUFT.
NUFT of a function $f(\x{1})$ is defined as:
\begin{equation}
\begin{aligned}
F(\x{2})=\nufourier{f(\x{1})}=\int \!\! \int f(\x{1}) \exp(- i \pi \x{1}\x{2}) d\x{1}.
\end{aligned}
\label{eqn:nufourier}
\end{equation}

Although the form of NUFT is similar to that of uniform Fourier transform, the coordinate $\x{1}$ is non-uniform-sampled and $\x{2}$ is uniform-sampled, unlike uniform Fourier transform.
For the numerical implementation of Eq.(\ref{eqn:nufourier}), it is necessary to use non-uniform fast Fourier transform (NUFFT) which has the complexity of $O(N \log N)$.
Many methods for NUFFT have been proposed over the course of the past twenty years or so \cite{nufft1,nufft2,nufft3,nufft4}.
NUFFTs are based on the combination of an interpolation and the uniform FFT.
In this paper, we used L. Greengard and J. Y. Lee's NUFFT \cite{nufft4}.
For more details, see \cite{nufft4}.

\section{Result}

\begin{figure}[htbp]
\centerline{
\includegraphics[width=12cm]{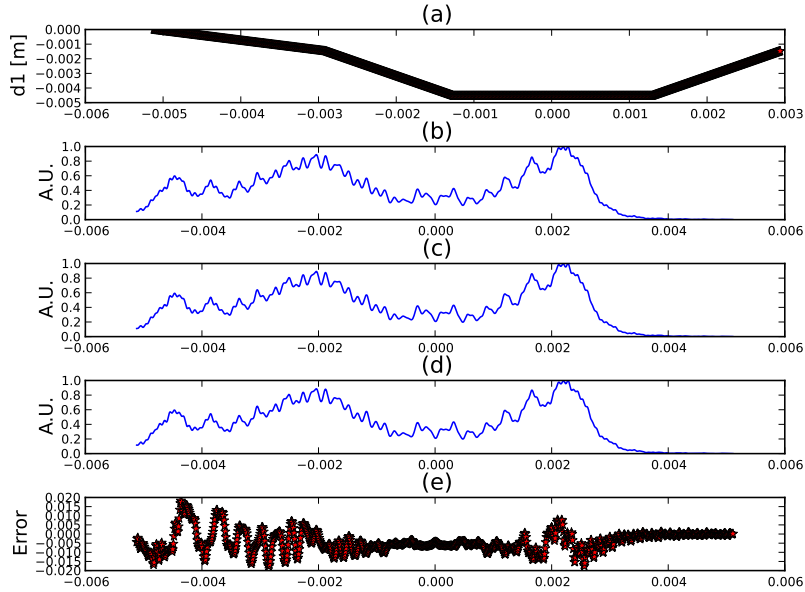}}
 \caption{Intensity profiles of diffraction results from a source surface composed of four small planar surfaces with 128 points, which are tilted $-30^{\circ}$, $-50^{\circ}$, $0^{\circ}$ and $+50^{\circ}$to $\x{1}$, respectively. (a) source surface (b) diffraction result by Eq.(\ref{eqn:huy_diff2}) (c) diffraction result by Eq.(\ref{eqn:fre_diff1}) (d) diffraction result by Eq.(\ref{eqn:fre_diff_nuf}) (e) Absolute error between (b) and (d).}
\label{fig:complex-planar}
\end{figure}

\noindent Let us examine our method using two source surfaces with arbitrary shapes in one dimension: a surface composed of four tilted planar surfaces, and quadratic curve.

We used the distance $z_0=1$ m, the wavelength of 633 nm, and the number of sampling points on source and destination $N=1,024$.
The sampling rates on the source and destination surface are $p=10 \mu$ m.
We used a planar wave as the incident wave that is perpendicular to the optical axis.

Figure \ref{fig:complex-planar} shows a source surface composed of four small planar surfaces with 128 points, which are tilted $-30^{\circ}$, $-50^{\circ}$, $0^{\circ}$ and $+50^{\circ}$ to $\x{1}$, respectively.
The horizontal axis in (a) indicates the position on $\x{1}$ in metric units.
The horizontal axes in (b)-(e) indicates the position on $\x{2}$ in metric units.
The sampling rates on these small planar surfaces are $p$, however, the sampling rates on $\x{1}$ are $|p \cos(-30^{\circ})|$, $|p \cos(-50^{\circ})|$, $|p \cos(0^{\circ})|$ and $|p \cos(+50^{\circ})|$, respectively.
The destination planar surface is not inclined to $\x{2}$.
Figures \ref{fig:complex-planar} (b)-(d) show the intensity profiles of the diffraction results by Eq.(\ref{eqn:huy_diff2}), Eq.(\ref{eqn:fre_diff1}) with direct integral and our method (Eq.(\ref{eqn:fre_diff_nuf})), respectively.
Figure \ref{fig:complex-planar} (e) depicts the absolute error between (Eq.(\ref{eqn:huy_diff2})) and Eq.(\ref{eqn:fre_diff_nuf}).
The absolute error falls into within approximately 0.025.

\begin{figure}[htbp]
\centerline{
\includegraphics[width=12cm]{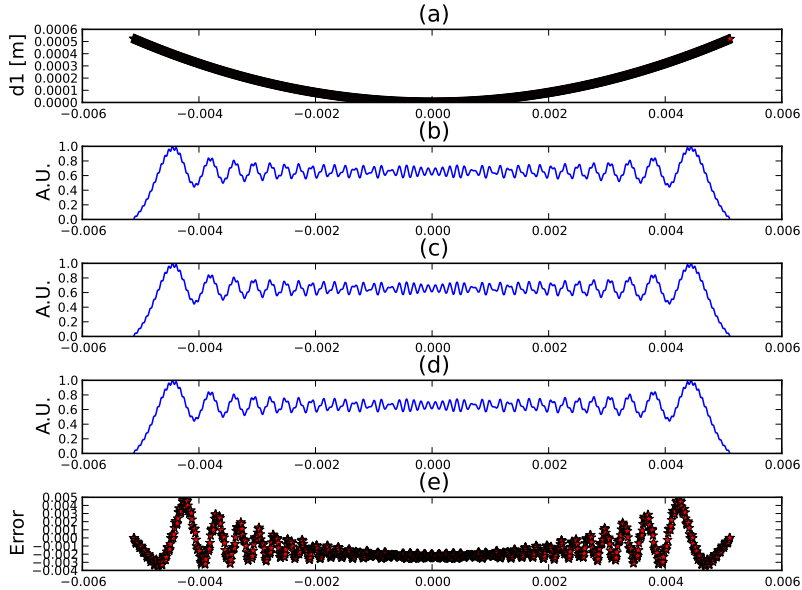}}
 \caption{Intensity profiles of diffraction results from quadratic curve surface. (a) source surface (b) diffraction result by Eq.(\ref{eqn:huy_diff2}) (c) diffraction result by Eq.(\ref{eqn:fre_diff1}) (d) diffraction result by Eq.(\ref{eqn:fre_diff_nuf}) (e) absolute error between (b) and (d).}
\label{fig:quad-curve}
\end{figure}

Figure \ref{fig:quad-curve} shows a source surface with quadratic curve.
The sampling rate on $\x{1}$ according to the source surface depends on the slope of the quadratic curve.
Figure \ref{fig:quad-curve} (b)-(d) shows the intensity profiles of the diffraction results by Eq.(\ref{eqn:huy_diff2}), Eq.(\ref{eqn:fre_diff1}) and our method (Eq.(\ref{eqn:fre_diff_nuf})), respectively.
Figure \ref{fig:quad-curve} (e) depicts the absolute error between Eq.(\ref{eqn:huy_diff2}) and Eq.(\ref{eqn:fre_diff_nuf}).
The absolute error falls into within approximately 0.005.
The primary factor of these absolute errors in Figs.\ref{fig:complex-planar} and \ref{fig:quad-curve} is the approximations by Eqs.(\ref{eqn:r2}) and (\ref{eqn:fre_diff4}).

\section{Conclusion}
\noindent We proposed Fresnel diffraction calculation from a source surface with arbitrary shape to a planar destination surface, using one NUFFT.
We have had only a direct integral for calculating such diffraction thus far.
Unfortunately, it is very time consuming because the calculation cost takes $O(N^2)$ in one dimensional or $O(N^4)$ in two dimensional cases.
In contrast, the calculation cost of our method is $O(N \log N)$ in one dimensional or $O(N^2 \log N)$ in two dimensional cases using NUFFT.
The method is very useful for calculating a CGH from a three-dimensional object composed of multiple polygons or arbitrary shape surfaces.
In our next work, we will show the fast calculation of a CGH from such 3D objects using this method.

\section*{Acknowledgments}
\noindent This work is supported by the Ministry of Internal Affairs and Communications, Strategic Information and Communications R\&D Promotion Programme (SCOPE), Japan Society for the Promotion of Science (JSPS) KAKENHI (Young Scientists (B) 23700103) 2011, and the NAKAJIMA FOUNDATION.
\end{document}